\documentstyle[12pt]{article}

\newcommand{\be}{\begin{equation}}
\newcommand{\ee}{\end{equation}}
\newcommand{\bea}{\begin{eqnarray}}
\newcommand{\eea}{\end{eqnarray}}

\newcommand{\EQ}{\begin{equation}}
\newcommand{\EN}{\end{equation}}

\renewcommand{\thefootnote}{\fnsymbol{footnote}}

\begin{document}
\begin{titlepage}
\rightline{DFTT 71/96}
\rightline{NORDITA 96/73 P}
\rightline{\hfill November 1996}

\vskip 2.2cm

\centerline{\Large \bf The field theory limit of multiloop string 
amplitudes}

\vskip 1.2cm

\centerline{\bf Paolo Di Vecchia\footnote{e-mail:
DIVECCHIA@nbivms.nbi.dk}, 
Lorenzo Magnea\footnote{On leave from Universit\`a di Torino, Italy}, 
Raffaele Marotta}
\centerline{\sl NORDITA}
\centerline{\sl Blegdamsvej 17, DK-2100 Copenhagen \O, Denmark}

\vskip .2cm

\centerline{\bf Alberto Lerda}
\centerline{\sl Dipartimento di Scienze e Tecnologie Avanzate and}
\centerline{\sl Dipartimento di Fisica Teorica, Universit\`a di Torino}
\centerline{\sl Via P.Giuria 1, I-10125 Torino, Italy}
\centerline{\sl and I.N.F.N., Sezione di Torino}

\vskip .2cm

\centerline{\bf Rodolfo Russo}
\centerline{\sl Dipartimento di Fisica, Politecnico di Torino}
\centerline{\sl Corso Duca degli Abruzzi 24, I-10129 Torino, Italy}
\centerline{\sl and I.N.F.N., Sezione di Torino}
\vskip 1cm

\begin{abstract}
We report on recent progress in the use of string techniques 
for the computation of field theory amplitudes. We show how one--loop
renormalization constants in Yang--Mills theory can be computed using
the open spinning string, we review the calculation of two--loop scalar
amplitudes with the bosonic string, and we briefly indicate how the
technique can be applied to the two--loop vacuum bubbles of 
Yang--Mills theory.
\end{abstract}

\end{titlepage}

\newpage
\renewcommand{\thefootnote}{\arabic{footnote}}
\setcounter{footnote}{0}
\setcounter{page}{1}
\section{Introduction}
\label{intro}

All string theories contain a parameter,
the string tension $T=1/(2\pi\alpha')$, having the dimension of 
a mass squared. In the infinite tension limit, $\alpha'\to 0$,
the heavy string states become infinitely massive
and decouple, while the light states survive. Thus, when
$\alpha'\to 0$, a string theory simply reduces
to a quantum field theory of pointlike objects.
Moreover, the parameter $1/\alpha'$ acts in string theory as
an ultraviolet regulator in the integrals over loop
momenta, making the string free from ultraviolet
divergences. For these reasons, string theory 
may be useful not only for the construction of unified theories
but also as an efficient tool to understand the structure of 
perturbative field theories, and to compute regularized amplitudes.
 
There are several features that make string theories very useful 
for this purpose. First of all, at each order of string perturbation 
theory, one does not get the large number of diagrams that one 
encounters in field theories. For example, with closed strings there
is only one diagram at each perturbative order. 
Secondly, there exist compact and explicit expressions for 
string scattering amplitudes that are valid at {\it any} 
perturbative order~\cite{copgroup}. 
This is to be contrasted with what happens
in field theory where no such all--loop formulas exist.
Finally, string amplitudes are always written in a way
that takes maximal advantage of all symmetries, and in
particular of gauge invariance.
All these properties have been known since the very
early days of string theory, but it
is only in more recent times that they have been 
exploited for explicit calculations in Yang--Mills theory~\cite{various}.

The aim of this talk is to review and extend some of the results of 
Refs.~\cite{oneloop} and~\cite{twoloop}, where the
limit $\alpha' \to 0$ of bosonic string theory was performed in order
to analyze the structure of ultraviolet divergences in 
Yang--Mills theory at one loop, and to compute scattering amplitudes
of a $\Phi^3$ theory at two loops.

Our analysis clearly shows that only some corners of the string
moduli space contribute to the field theory limit, namely those
corners where the integrand of the string amplitude exhibits a
singular behaviour when $\alpha' \to 0$. 
For each such corner of moduli space, the string gives directly a
Schwinger parameter integral, with momentum integrations, Lorentz and 
gauge algebra already performed. As a consequence, different regularizations
are easily implemented: we work mainly with dimensional regularization,
but it would be possible, for example, to use a proper--time cutoff on the
Schwinger parameters. In particular, string theory gives correctly also
the contributions of the diagrams (such as tadpoles or vacuum bubbles)
that vanish in dimensional regularization. 

It is also worth pointing out that actually there are several ways to 
take the infinite tension limit in a string amplitude. 
One of these, perhaps the most obvious and natural one, is to
let $\alpha' \to 0$  in such a way that only the massless states 
survive. Performing this limit in the bosonic string, for example, one 
recovers a non--abelian Yang--Mills theory from the open sector,
unified with an extended version of gravity from the closed sector, 
containing besides the graviton also an antisymmetric tensor and a dilaton.
However, there are other consistent possibilities.
For example, in the open bosonic string one can take the infinite
tension limit and retain only the massive scalar state instead of the
massless vector ones. This can be achieved by giving an arbitrary mass 
$m$ to the scalars, such that  $\alpha'\,m^2=-1$ when $\alpha' \to 0$, and
by suitably modifying the measure of integration in the string amplitudes 
to avoid the inconsistencies associated with tachyon
propagation. Thus, in this case, the infinite tension limit of the open
bosonic string yields a $\Phi^3$ field theory~\cite{scherk}. 

The talk is organized as follows: in Sect. 2 we show how to compute 
renormalization constants in Yang--Mills theory at one loop, starting from 
the two--gluon scattering amplitude of the open spinning string.
This is an extension of the results of Ref.~\cite{oneloop},
where similar calculations were performed in the open 
bosonic string and the inconsistencies due to tachyon
exchanges were removed by hand. One sees that the results are independent 
of the string model one starts with. In Sect. 3 we show how to 
obtain the correctly normalized two--loop vacuum bubbles of the
$\Phi^3$ theory starting from the two--loop tachyon amplitudes
of the open bosonic string. Finally, we conclude by briefly indicating how 
the same calculation can be generalized to the vacuum bubbles of Yang--Mills 
theory.

\section{Yang--Mills Renormalization Constants \\
from the Spinning String}
\label{nsr}

Given two gluons with momenta $p_1$ and $p_2$, polarizations 
$\varepsilon_1$ and $\varepsilon_2$, and $SU(N)$ color indices $a_1$ 
and $a_2$, the corresponding scattering amplitude at one loop in the 
open spinning string is 
\be
A^{(1)}_2 = N \, {\mathrm Tr}\left(\lambda^{a_1}
\lambda^{a_2}\right)~ 
\varepsilon_1\cdot\varepsilon_2~p_1\cdot p_2 ~\frac{{g_d}^2}{(4\pi)^{d/2}}
\,(2\alpha')^{2-d/2}~R\left(p_1\cdot p_2\right)~~~,
\label{twoglu}
\ee
where, with the same normalizations of Ref.~\cite{oneloop},
$g_d$ is the (dimensionful) Yang--Mills coupling constant in $d$ 
dimensions, and $R(s)$ is defined by
\bea
R(s) &=&\sum_{\mathbf{a}}
\int_0^1 \frac{dk}{k^{3/2}} \left(-\frac{1}{2}\log k\right)^{-d/2}
\prod_{n=1}^\infty \left(1-k^n\right)^{2-d}\,
Z_{\mathrm F}^{[\mathbf{a}]}(k)
\nonumber \\
&\times&\int_k^1dz_2  
\Big[
\big(\partial_2 G_{\mathrm B}(1,z_2)\big)^2
-\big(G_{\mathrm F}^{[\mathbf{a}]}(1,z_2)\big)^2
\Big]~{\mathrm e}^{2\alpha' s \,G_{\mathrm B}(1,z_2)}~~~.
\label{2gluon}
\eea
In this equation, $\sum_{\mathbf{a}}$ denotes the sum over the even 
spin structures, $k$ is the modular parameter of the annulus,
$Z_{\mathrm F}^{[\mathbf{a}]}(k)$ is the fermionic partition function 
of the $\mathbf{a}$--th spin structure, and $G_{\mathrm B}$ and 
$G_{\mathrm F}^{[\mathbf{a}]}$ are the bosonic and fermionic Green 
functions, respectively. 
Notice that if the gluons are on mass shell, the two--gluon amplitude 
in Eq. \ref{twoglu} becomes ill--defined, because the kinematical prefactor
vanishes, while the integral $R$ diverges. As explained in
Ref.~\cite{oneloop}, this problem can be consistently cured 
by keeping the gluons off--shell. 

If we are interested in pure Yang--Mills theory, the sum in 
Eq. \ref{2gluon} must be restricited to the two even spin structures 
(${\mathbf{a}}=0,1$) of the Neveu--Schwarz sector, which describe
bosonic string states circulating in the loop\footnote{The third even 
spin structure corresponds to fermionic states of the Ramond sector
propagating in the loop, and should be considered if one were interested 
in supersymmetric Yang--Mills theories.}. The corresponding fermionic 
partition functions are
\be
Z_{\mathrm F}^{[\mathbf{a}]}(k) = \frac{(-1)^{\mathbf{a}}}{2}~
\prod_{n=1}^\infty \left(1+(-1)^{\mathbf{a}}k^{n-1/2}\right)^{2-d}~~~,
\label{part}
\ee
while the fermionic Green functions are
\be
G_{\mathrm F}^{[\mathbf{a}]}(z,w) = \sum_{n=-\infty}^\infty
(-1)^{({\mathbf{a}}+1)n} \frac{k^{n/2}}{k^nz-w}~~~,
\label{fermgre}
\ee
for ${\mathbf{a}}=0,1$. The bosonic Green function $G_{\mathrm B}$ is 
obviously independent of the spin structure and is given by
\be
G_{\mathrm B}(z,w) = 
\frac{1}{2 \log k} 
\left( \log \frac{z}{w} \right)^2
+ \log \left[\frac{z - w}{\sqrt{z\,w}}\prod_{n = 1}^\infty 
\frac{\left(1 - k^n \frac{w}{z} \right)
\left(1 - k^n \frac{z}{w} \right)}{\left(1 - k^n \right)^2} \right] 
\label{bosgre}
\ee
To discuss the field theory limit of Eq. \ref{twoglu} it is convenient
to introduce the variables $\tau = -\log k/2$ and $\nu = -\log z_2/2$,
which turn out to be related directly to the
Schwinger proper times $t$ and $t_1$ of the Feynman diagrams 
contributing to the two--point function~\cite{oneloop}. In particular, 
$t \sim \alpha' \tau$ and $t_1 \sim \alpha' \nu$, 
where $t_1$ is the proper time associated with one of the two internal 
gluon propagators, while $t$ is the total proper time around the loop. 
In the field theory limit these proper times have to remain finite, and 
thus the limit $\alpha' \to 0$ must be accompanied by the limit 
$\{ \tau, \nu \} \to \infty$ in the integrand of Eq. \ref{2gluon}.
Thus the field theory limit of a string amplitude is determined by
the asymptotic behavior of the Green functions $G_{\mathrm B}$ and
$G_{\mathrm F}^{[\mathbf{a}]}$ for large $\tau$, which is given by
\be
G_{\mathrm B}(\nu,\tau) \sim \nu -\frac{\nu^2}{\tau} 
- {\mathrm e}^{-2\nu}
-{\mathrm e}^{-2\tau+2\nu} + 2{\mathrm e}^{-2\tau}~~~,
\label{gbtau}
\ee
and
\be
G_{\mathrm F}^{[\mathbf{a}]}(\nu,\tau) \sim  1
+(-1)^{\mathbf{a}}{\mathrm e}^{-\tau+2\nu}+
(-1)^{{\mathbf{a}+1}}{\mathrm e}^{-\tau}~~~.
\label{gftau}
\ee
We now substitute these results in Eq. \ref{2gluon} and keep only
those terms that remain finite when $k={\mathrm e}^{-2\tau} \to 0$.
While in the bosonic string tachyon exchanges produce divergent terms 
which must be discarded by hand~\cite{oneloop}, here our procedure leads 
directly to the desired contributions due to gluon exchanges, since in the 
spinning string tachyons are projected out by the sum over spin structures. 
Indeed, by defining ${\hat \nu}\equiv\nu/\tau$ and performing simple 
manipulations, we get
\be
R(s) =\int_0^\infty d\tau \int_0^1 d{\hat \nu} ~\tau^{1-d/2}\,
{\mathrm e}^{2\alpha'\,s\,({\hat\nu}-{\hat\nu}^2)\tau} 
\, \left[(1-2{\hat\nu})^2(d-2)-8\right]~~~.
\label{limRint}
\ee
which is precisely the same result that emerges from the bosonic 
string~\cite{oneloop} after discarding tachyons. In this case,
the term proportional to $(d-2)$ in the square bracket is generated by the 
bosonic Green function of Eq. \ref{2gluon}, while the factor $-8$ is 
produced by the fermionic Green functions. The integrals in Eq. \ref{limRint}
are now elementary and yield
\be 
R(s) = - \Gamma\left(2-\frac{d}{2}\right)\, (- 2 \alpha' s)^{d/2-2}~
\frac{6-7d}{1-d}~B\left(\frac{d}{2}-1,\frac{d}{2}-1\right)~~~,
\label{Rfin}
\ee
where $B$ is the Euler beta function.

If we substitute Eq. \ref{Rfin} into Eq. \ref{twoglu}, we see that 
the $\alpha'$ dependence cancels, as it must. 
The ultraviolet finite string amplitude in Eq. \ref{twoglu} 
has been replaced by a field theory amplitude which diverges in four 
space--time dimensions, because of the pole in the $\Gamma$ function in 
Eq. \ref{Rfin}. The result coincides exactly with the one--loop gluon vacuum
polarization of the $SU(N)$ gauge field theory that one computes
with the background field method, in the Feynman gauge and using dimensional 
regularization.

Setting $d=4-2\epsilon$ and defining as usual a dimensionless coupling 
constant $g = g_d\,\mu^{-\epsilon}$, with $\mu$ an arbitrary mass scale,
the divergent part of the two--point amplitude is 
\be
A^{(1)}_2 = - N \,\frac{g^2}{(4\pi)^2}\,\frac{11}{3} 
\,\frac{1}{\epsilon}~
\delta^{ab}\,\varepsilon_1\cdot
\varepsilon_2~p_1\cdot p_2
+ O(\epsilon^0)~~~.
\label{twofin}
\ee
 From this result it is immediate to extract the minimal subtraction wave 
function renormalization constant at one loop,
\be
Z_A= 1 + N\,\frac{g^2}{(4\pi)^2}\,\frac{11}{3}\,\frac{1}{\epsilon}~~~.
\label{za}
\ee
With similar techniques, one can compute the three and four--point
vertex renormalization constants, $Z_3$ and $Z_4$, at one loop
and verify that they satisfy the correct Ward identity
$Z_3=Z_4=Z_A$ appropriate to the background field method~\cite{oneloop}.

\section{Two--Loop Diagrams in $\Phi^3$ Theory}
\label{phi}

We now consider the open bosonic string and outline the procedure
to obtain the two--loop Feynman diagrams of the $\Phi^3$ field theory
from the two--loop string amplitudes in the limit $\alpha'\to 0$.
This is a preliminary but important step towards the more
interesting case of multiloop amplitudes in Yang--Mills
theories. By selecting scalar particles instead of
gluons, one can avoid the computational difficulties related to the 
fact that the gluons are not the lowest states of the spectrum, and 
one can focus on the precise identification of the corners of moduli 
space contributing to the field theory limit.

Our starting point is the planar two--loop scattering amplitude of $M$ 
scalar particles of momenta $p_1,\ldots,p_M$, and $U(N)$ color indices
$a_1,\ldots,a_M$, which is 
\bea
A^{(2)}_M & = & N^2\,{\mathrm Tr}(\lambda^{a_1}
\cdots \lambda^{a_M})\,
\frac{1}{(4\pi)^d}\,\frac{g^{2+M}}{2^{8+3M}}\,
(2 \alpha' )^{3-d+M} \nonumber \\
& \times & \int \frac{ dk_1}{k_{1}^{2}} \int \frac{ d k_2}{k_{2}^{2}}
\int \frac{ d \eta_1 }{ ( 1 -  \eta_1 )^{2} }
\prod_{i=1}^{M} \int \left[ \frac{d z_i}{V_{i} ' (0) } \right] 
\label{2Mtac}\\ & \times &
\left[\frac{1}{4} \left(\log k_1 \log k_2 - \log^2 \eta_1 \right)
\right]^{-d/2} \prod_{i<j} \left[{{\exp\left({\mathcal G}^{(2)}
(z_i,z_j)\right)}
\over{\sqrt{V'_i(0)\,V'_j(0)}}}\right]^{2\alpha' p_i\cdot p_j}~~~.
\nonumber
\eea
In this formula ${\mathcal G}^{(2)}(z_i,z_j)$ is the two--loop
bosonic Green function, while $V_i(z)$ parametrizes the local
coordinates around the puncture $z_i$ ~\cite{scho}. Notice that
in Eq. \ref{2Mtac} we have used the two--loop integration
measure of the open bosonic string~\cite{meas} in the limit
of small $k_1$ and $k_2$. This is the obvious two--loop generalization
of the limit $k={\mathrm e}^{-2\tau}\to 0$ considered in the 
one loop case.

Eq. \ref{2Mtac} serves as a `master formula' for all two--loop amplitudes
of the theory defined by the Lagrangian
\be
L = {\mathrm Tr} \left[ \partial_{\mu} \Phi 
\partial^{\mu} \Phi + m^2 \Phi^2
- \frac{g}{3!} \Phi^3  \right]~~,
\label{Phi3lag}
\ee
where $\Phi = \Phi^a\lambda^a$ is a scalar field in the adjoint 
representation of $U(N)$. All amplitudes are determined by the single 
function ${\mathcal G}^{(2)}(z_i,z_j)$, which must be evaluated in 
different corners of moduli space. The integrations in Eq. \ref{2Mtac} 
correspond to a sum over all the different positions of the vertex operators 
representing external legs, as well as over all the different shapes of 
the world--sheet surface. In our case the surface is represented in the 
Schottky parametrization, as shown in Fig. 1.

\vspace{1cm}
\unitlength .8mm
\linethickness{0.4pt}
\begin{picture}(125.00,40.00)(10,75)
\put(80.00,100.00){\circle{8.67}}
\put(100.00,100.00){\circle{12.00}}
\put(126.00,100.00){\circle{12.00}}
\multiput(150.00,120.00)(0.11,-0.61){4}{\line(0,-1){0.61}}
\multiput(150.46,117.56)(0.10,-0.61){4}{\line(0,-1){0.61}}
\multiput(150.86,115.12)(0.11,-0.81){3}{\line(0,-1){0.81}}
\multiput(151.20,112.68)(0.09,-0.81){3}{\line(0,-1){0.81}}
\multiput(151.48,110.24)(0.11,-1.22){2}{\line(0,-1){1.22}}
\multiput(151.70,107.80)(0.08,-1.22){2}{\line(0,-1){1.22}}
\put(151.86,105.37){\line(0,-1){2.44}}
\put(151.96,102.93){\line(0,-1){2.44}}
\put(152.00,100.49){\line(0,-1){2.44}}
\put(151.98,98.05){\line(0,-1){2.44}}
\multiput(151.90,95.61)(-0.07,-1.22){2}{\line(0,-1){1.22}}
\multiput(151.77,93.17)(-0.10,-1.22){2}{\line(0,-1){1.22}}
\multiput(151.57,90.73)(-0.09,-0.81){3}{\line(0,-1){0.81}}
\multiput(151.31,88.29)(-0.11,-0.81){3}{\line(0,-1){0.81}}
\multiput(151.00,85.85)(-0.09,-0.61){4}{\line(0,-1){0.61}}
\multiput(150.62,83.41)(-0.10,-0.57){6}{\line(0,-1){0.57}}
\put(155.00,100.00){\line(-1,0){150.00}}
\multiput(10.00,80.00)(-0.11,0.61){4}{\line(0,1){0.61}}
\multiput(9.58,82.44)(-0.09,0.61){4}{\line(0,1){0.61}}
\multiput(9.21,84.88)(-0.11,0.81){3}{\line(0,1){0.81}}
\multiput(8.89,87.32)(-0.09,0.81){3}{\line(0,1){0.81}}
\multiput(8.63,89.76)(-0.11,1.22){2}{\line(0,1){1.22}}
\multiput(8.41,92.20)(-0.08,1.22){2}{\line(0,1){1.22}}
\put(8.25,94.63){\line(0,1){2.44}}
\put(8.15,97.07){\line(0,1){2.44}}
\put(8.09,99.51){\line(0,1){2.44}}
\put(8.08,101.95){\line(0,1){2.44}}
\put(8.13,104.39){\line(0,1){2.44}}
\multiput(8.23,106.83)(0.08,1.22){2}{\line(0,1){1.22}}
\multiput(8.38,109.27)(0.10,1.22){2}{\line(0,1){1.22}}
\multiput(8.59,111.71)(0.09,0.81){3}{\line(0,1){0.81}}
\multiput(8.84,114.15)(0.10,0.81){3}{\line(0,1){0.81}}
\multiput(9.15,116.59)(0.10,0.68){5}{\line(0,1){0.68}}
\put(149.17,102.00){\makebox(0,0)[cc]{$B'$}}
\put(11.50,102.00){\makebox(0,0)[cc]{$A'$}}
\put(73.60,102.00){\makebox(0,0)[cc]{$A$}}
\put(86.20,102.00){\makebox(0,0)[cc]{$B$}}
\put(92.00,102.00){\makebox(0,0)[cc]{$C$}}
\put(108.67,102.00){\makebox(0,0)[cc]{$D$}}
\put(117.50,102.00){\makebox(0,0)[cc]{$D'$}}
\put(135.20,102.00){\makebox(0,0)[cc]{$C'$}}
\put(80,108.00){\makebox(0,0)[cc]{${\cal K}_2$}}
\put(100.70,110.00){\makebox(0,0)[cc]{${\cal K}_1$}}
\put(126.70,110.00){\makebox(0,0)[cc]{${\cal K}_1'$}}
\end{picture}

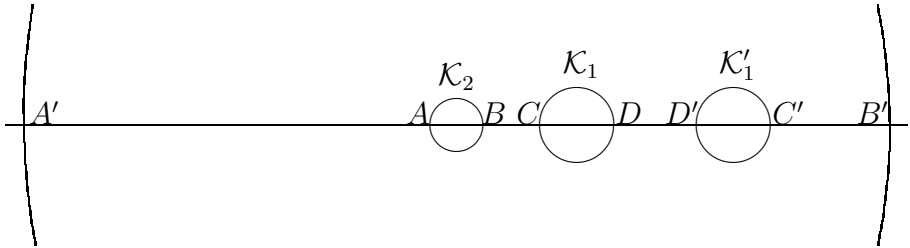
Fig. 1: In the Schottky parametrization, the two--annulus corresponds 
to the part of the upper--half plane which is inside the big
circle passing through $A'$ and $B'$, and which is outside the
circles ${\cal K}_1$, ${\cal K}_1'$ and ${\cal K}_2$.
\vspace{.5cm}

The width of the two--annulus around the hole represented by
the two circles ${\cal K}_1 \sim {\cal K}_1'$ is proportional to $k_1$,
while the width around the other hole is proportional to $k_2$; 
on the other hand $\eta_1$ can be seen as the ``distance'' between the 
two loops. The points $A$, $B$, $C$ and $D$ have to be identified with
$A'$, $B'$, $C'$ and $D'$ respectively, so that it is easy to realize 
that the two segments $(AA')$ and $(DD')$ represent the two inner 
boundaries of the two--annulus, while the union of $(BC)$ and $(C'B')$ 
represents the external boundary. 

In the limit $\alpha'\rightarrow 0$ the integrals in Eq. \ref{2Mtac} are 
dominated by the thin two--annulus with $k_1$,$k_2 \rightarrow 0$; 
moreover if $\eta_1 \rightarrow 0$, the distance between the two holes 
vanishes and the loops glue together giving 1PI diagrams, while if $\eta_1
\rightarrow 1$, one recovers the reducible diagrams in which the loops are 
separated by a propagator. Unlike what we did in the previous
section, we want now to select the scalar state; the easiest way to 
do this is to relax the condition $1 + \alpha' m^2 =0$ for the mass of the
scalar state and change it into $a + \alpha' m^2 =0$. We can now transform 
the tachyon into a normal scalar particle with arbitrary mass $m$ by 
rewriting each double pole in the measure in Eq.~(\ref{2Mtac}) according to
\be
x^{-2} \rightarrow x^{-1-a}=x^{-1}\exp \left[-a \log x \right]=
x^{-1}\exp \left[m^2 \alpha' \log x \right] ~~.
\label{mass}
\ee

As an easy example of the procedure just outlined, we will calculate the
vacuum bubble diagrams. In this simple case there is no dependence on 
the Green function and we only have to integrate over the Schottky 
parameters. However it is more convenient to introduce new 
variables
\be
q_1={k_2\over\eta_1}~,~~~q_2={k_1\over\eta_1}~,~~~q_3=\eta_1~~,
\label{qvariable}
\ee
so that Eq. \ref{2Mtac} becomes
\bea
A^{(2)}_0 & = &  \frac{N^3}{(4\pi)^d}\,\frac{g^{2}}{2^{8}}\,
(2 \alpha' )^{3-d}
\!
\int_0^{1} \frac{d q_3 }{ (1 - q_3)^{1+a}~ q_3^{1+a}}
\int_{0}^{q_3}\frac{dq_2}{q_{2}^{1+a}} \int_{0}^{q_2}
\frac{d q_1 }{ q_1^{1+a}}
\nonumber\\ & \times & \left[\frac{1}{4}\left(\log q_1 \log
q_2 + \log q_1 \log q_3 + \log q_2 \log q_3 \right)
\right]^{-d/2} \label{0point}~~.
\eea
The region of integration we used for the $q_i$ can be derived 
from the geometrical interpretation described above~\cite{oneloop}.
We now consider the reducible and irreducible diagrams separately.
In the first case ($q_3\rightarrow 1$), Eq. \ref{0point} becomes
\be
A^{(2)}_0\Big|_{\mathrm red} 
= \frac{N^3}{(4\pi)^d}\,\frac{g^{2}}{2^{5}}\,
\int_0^{\infty}dt_3 \int_0^{\infty}dt_2 \int_0^{t_2}dt_1~
{\mathrm e}^{-m^2(t_1+t_2+t_3)}~(t_1\,t_2)^{-d/2}~~,\label{vacuumr}
\ee
where we introduced the mass $m$ as explained in Eq. \ref{mass}, and the
Schwinger proper times $t_i$ according to
\be
t_1=-\alpha'\log q_1\,,~~~t_2
=-\alpha'\log q_2\,,~~~t_3=-\alpha'\log (1-q_3)~~;
\label{ti}
\ee
in the second case ($q_3\rightarrow 0$), Eq. \ref{0point} becomes
\bea
A^{(2)}_0\Big|_{\mathrm irr} & = &  
\frac{N^3}{(4\pi)^d}\,\frac{g^{2}}{2^{5}}
\!
\int_0^{\infty}dt_3 \int_0^{t_3}dt_2 \int_0^{t_2}dt_1~
e^{-m^2(t_1+t_2+t_3)}\\
&\times& (t_1t_2+t_1t_3+t_2t_3)^{-d/2}~~;
\nonumber
\eea
where $t_3$ is now defined by $t_3=-\alpha'\log q_3$.

One can check that the results above coincide with the Schwinger
parametrization of the same diagrams in field theory. We have 
explicitly verified that this method gives the correct results
also in the presence of external states, even if they are off shell; in 
this latter case the result depends on the particular choice of the 
local coordinates $V_i$. It turns out that the correct field
theory amplitudes are obtained if one defines $V_i$ to satisfy
\be
{\left( V_{i} ' (0) \right)}^{-1} = \left\vert{1\over z_i - \rho_a} -
{1\over z_i - \rho_b}\right\vert~~, \label{Vi}
\ee
where $\rho_a$ and $\rho_b$ are the two fixed points on the left and on 
the right of $z_i$ (this definition of $V_i$ corresponds to the one used in
Ref.~\cite{kajsato}).

\section{Conclusions}
\label{conc}

We would like to conclude by noting that the procedure described in 
Section 3 can be readily generalized to gluon propagation, and in fact we
have already completed the calculation of the vacuum bubble diagrams in 
Yang--Mills theory at two loops, obtaining the correct result. Lack of space
prevents us from describing the calculation in this contribution. However 
we note the main features: it is clearly necessary to expand the various 
terms in the string--theoretic measure to next--to--leading order in the 
multipliers $k_i$. Then tachyon poles are discarded as was done at one loop. 
The main new feature is the diagram with the four gluon vertex, which can 
be traced to a contact term left over from the propagation of a tachyon in 
the corresponding channel. However it is interesting to notice that there
does not appear to be a natural mapping between individual diagrams
and points in moduli space. Only when the different contributions are 
recombined the correct answer is recovered in the form of a standard
Schwinger parameter integral~\cite{next}. 

\vskip 1.0cm
{\large {\bf Acknowledgements}}

Work supported by the European Commission TMR programme 
ERBFMRX-CT96-0045, in which R.R. is associated to 
University of Torino.


\vskip 0.5cm



\begin{thebibliography}{99}

\newcommand{\Journal}[4]{{#1} {\bf #2}, #3 (#4)}
\newcommand{\NCA}{\em Nuovo Cimento}
\newcommand{\NIM}{\em Nucl. Instrum. Methods}
\newcommand{\NIMA}{{\em Nucl. Instrum. Methods} A}
\newcommand{\NPB}{{\em Nucl. Phys.} B}
\newcommand{\PLB}{{\em Phys. Lett.}  B}
\newcommand{\PRL}{\em Phys. Rev. Lett.}
\newcommand{\PRD}{{\em Phys. Rev.} D}
\newcommand{\ZPC}{{\em Z. Phys.} C}



\bibitem{copgroup} See, for example, P. Di Vecchia, {\it ``Multiloop  
amplitudes in string theory''} in Erice, {\it Theor. Phys.} (1992), 
and references therein.

\bibitem{various} V. S. Kaplunovsky, \Journal{\NPB}{307}{145}{1988}, 
\Journal{\NPB}{382}{436}{1992}.
Z. Bern and D. A. Kosower, \Journal{\PRD}{38}{1888}{1988};
\Journal{\NPB}{321}{605}{1989}; 
\Journal{\NPB}{379}{451}{1992}. 
Z. Bern and D. C. Dunbar, \Journal{\NPB}{379}{562}{1992}.
Z. Bern, D. C. Dunbar and T. Shimada, \Journal{\PLB}{312}{277}{1993},
{\tt hep-th/9307001}.
Z. Bern, L. Dixon and D. A. Kosower, \Journal{\PRL}{70}{2677}{1993},
{\tt hep-ph/9302280}.
K. Roland, \Journal{\PLB}{289}{148}{1992}.

\bibitem{oneloop} P. Di Vecchia, A. Lerda, L. Magnea and R. Marotta,
\Journal{\PLB}{351}{445}{1995}, {\tt hep-th/9502156}.
P. Di Vecchia, A. Lerda, L. Magnea, R. Marotta and R. Russo, 
\Journal{\NPB}{469}{235}{1996}, {\tt hep-th/9601143};
in Erice, {\it Theor. Phys.} (1995), {\tt  hep-th/9602055};
proceedings of the 29th Ahrenshoop Symposium, Buckow, Germany,
(August 1995), {\tt hep-th/9602056}.

\bibitem{twoloop} P. Di Vecchia, A. Lerda, L. Magnea, R. Marotta and 
R. Russo, \Journal{\PLB}{388}{65}{1996}, {\tt hep-th/9607141}.

\bibitem{scherk}J. Scherk, \Journal{\NPB}{31}{222}{1971}.

\bibitem{scho} P. Di Vecchia, F. Pezzella, 
M. Frau, K. Hornfeck, A. Lerda and S. Sciuto, 
\Journal{\NPB}{322}{317}{1989}.

\bibitem{meas} P. Di Vecchia, M. Frau, A. Lerda and S. Sciuto, 
\Journal{\PLB}{199}{49}{1987}.

\bibitem{kajsato} K. Roland and H.-T. Sato, Niels Bohr Inst.
preprint {\bf NBI-HE 96 19},
April 1996, {\tt hep-th/9604152}.

\bibitem{next} P. Di Vecchia, A. Lerda, L. Magnea, R. Marotta and R. Russo,
in preparation.

\end{thebibliography}
\end{document}